\documentclass[epj,twocolumn]{webofc}
\usepackage[varg]{txfonts}   
\usepackage{textcomp}
\woctitle{ISVHECRI 2016}
\begin{document}
\title{The VHE anomaly in blazar spectra and related phenomena}
%
%

\author{Timur Dzhatdoev\inst{1}\fnsep\thanks{\email{timur1606@gmail.com}} \and
        Emil Khalikov\inst{1}\fnsep\thanks{\email{nanti93@mail.ru}} \and
        Anna Kircheva\inst{2} \and
        Alexander Lyukshin\inst{2}
}

\institute{Federal State Budget Educational Institution of Higher Education M.V. Lomonosov Moscow State University, Skobeltsyn Institute of Nuclear Physics (SINP MSU), 1(2), Leninskie gory, GSP-1, Moscow 119991, Russian Federation
\and
           Federal State Budget Educational Institution of Higher Education M.V. Lomonosov Moscow State University, Department of Physics, 1(2), Leninskie gory, GSP-1, Moscow 119991, Russian Federation
          }

\abstract{Most of the recent research on extragalactic $\gamma$-ray propagation focused on the study of the $\gamma \gamma \rightarrow e^{+} e^{-}$ absorption process (``absorption-only model''). Starting from a possible anomaly at very high energies (VHE, $E>$100 $GeV$), we briefly review several existing deviations from this model. The exotic interpretation of the VHE anomaly is not supported by the recent works. On the other hand, the process of intergalactic electromagnetic cascade development naturally explains these effects. We discuss phenomenology of intergalactic cascades and the main spectral signatures of the electromagnetic cascade model. We also briefly consider the hadronic cascade model; it also may explain the data, but requires low strength of magnetic field around the source of primary protons or nuclei.}
\maketitle
\section{Introduction}
\label{intro}

Observations made with imaging Cherenkov telescopes in the VHE range provide an opportunity to test extragalactic gamma-ray propagation models (e.g. \cite{J1,J2}). The majority of registered extragalactic VHE $\gamma$-ray sources are blazars --- active galactic nuclei that are likely to have narrowly collimated jets (with opening angles about 1$^{\circ}$-5$^{\circ}$) pointing  towards the observer.

VHE $\gamma$-rays are subject to the $\gamma \gamma \rightarrow e^{+} e^{-}$ absorption process. Some works found that the observed intensity in the optical depth region $\tau_{\gamma\gamma}>2$ is too high to be explained in the conventional framework of the ``absorption-only model'' that only takes into account  absorption of primary photons and their adiabatic losses \cite{J2,J3}. The statistical significance of the anomaly claimed in \cite{J2} amounts to 4.2 $\sigma$. The minimal energy at which the anomaly reveals itself shows a good correlation with the redshift of the source, indicating that the anomaly is not connected with intrinsic properties of the source but is the signature of the incompleteness of the absorption-only model.

The existence of the anomaly was initially interpreted as  evidence for $\gamma$--axion-like particle ($\gamma \rightarrow ALP$) oscillations \cite{J2}. This model is briefly considered in section~\ref{sec-1} of this paper. However, there are some other and less exotic ways to account for this anomaly. These explanations include questioning the validity of the theoretical EBL models invoked to explain observations. While the change of the EBL model could decrease the statistical significance of the anomaly at high energies, it certainly does not explain the whole observation set (see section~\ref{sec-2}). The anomaly can also be explained in the framework of the electromagnetic (section~\ref{sec-3}) or hadronic (section~\ref{sec-4}) intergalactic cascade models.

\section{Axion-like particles model}
\label{sec-1}

Axion-like particles (ALPs) are light bosons with zero spin that are a generalization of the axion --- a particle that served as a natural solution to the CP-symmetry violation problem \cite{J4,J5,J6}. Both particles are characterized by the two-photon coupling. However, unlike axions, ALPs do not have a connection between their two main parameters: their mass, $m_{a}$, and the two-photon coupling constant, $g_{a\gamma}$. Photons can oscillate into ALPs and back into photons in a similar way as neutrinos do. Indeed, if the $\gamma$--ALP mixing does occur, then the primary $\gamma$-rays can convert to ALPs near the source, thus avoiding absorption on the way to the observer, and then reconvert back into photons near the observer. These extra photons may account for the excess in intensity in the observable spectrum.

\begin{figure}
\centering
\includegraphics[width=8.0cm]{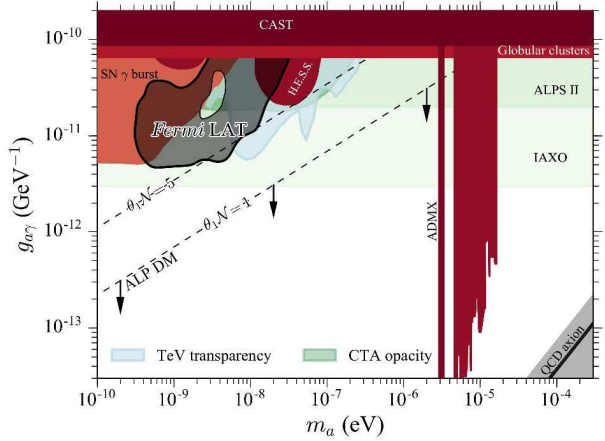}
\caption{Some constraints on the $\gamma$-ALP mixing parameters $g_{a\gamma}$ and $m_{a}$ (a figure from \cite{J12}).}
\label{fig-1}       
\end{figure}

\begin{figure}
\centering
\includegraphics[width=8.0cm]{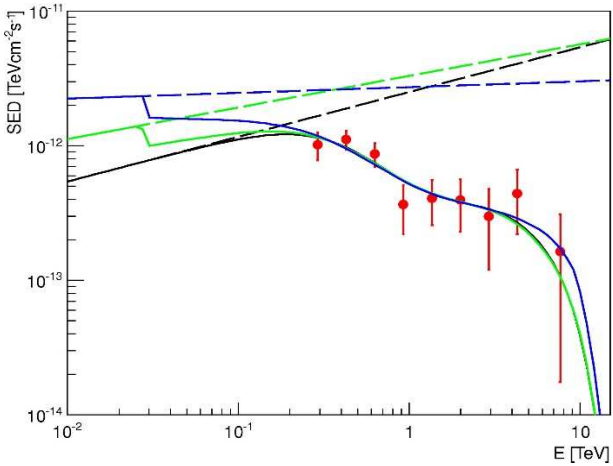}
\caption{Spectral signatures of $\gamma$-axion-like particle (ALP) mixing (a figure from \cite{J11}). Black curve denotes absorption-only model spectrum, green — relatively weak $\gamma$-ALP mixing, blue —  relatively strong $\gamma$-ALP mixing. Red circles with bars denote the VERITAS measurements together with uncertainties \cite{J9}.}
\label{fig-2}       
\end{figure}

The upper limit on $g_{a\gamma}$, that was established in \cite{J7}, is shown in fig. \ref{fig-1} as the brown area. Assuming that the VHE anomaly in blazar spectra is due to photon-ALP mixing, the lower limit on $g_{a\gamma}$ (depending on $m_{a}$) was first put in \cite{J8} (light blue area in fig.~\ref{fig-1}). Besides the aforementioned one, the $\gamma$-ALP mixing phenomenon has another distinct and readily identifiable signature that is shown in fig.~\ref{fig-2} for the case of blazar 1ES 0229+200 (redshift $z$= 0.14). Together with the spectral energy distribution (SED=$E^{2}dN/dE$) obtained with the VERITAS Cherenkov telescope \cite{J9}, fig.~\ref{fig-2} presents several model fits to this SED ($\gamma$-ALP mixing effects were taken into account according to \cite{J10}, see \cite{J11} for more details). $\gamma$-ALP mixing is effective above a certain critical energy, therefore a step-like irregularity in the shape of the observed spectrum is expected. The drop in intensity is usually about 1/3 of its original value (this result is valid when two photon polarization states attain equipartition with one polarization state of ALP above the critical energy). However, this signature of $\gamma$-ALP mixing was not found \cite{J12}, and the scenario in which $\gamma$-ALP oscillation can modify the $\gamma$-ray opacity of the Universe was strongly constrained (the corresponding excluded region of the $\gamma$-ALP mixing parameters is shown as gray area in the upper-left part of fig.~\ref{fig-1}).

\section{Electromagnetic cascade in the expanding magnetized Universe}
\label{sec-2}

In this work we also consider some other explanations of the anomaly, that do not require any new physics, but nevertheless go beyond the absorption-only model, introducing qualitatively new observable effects. In these models the primary particle (a $\gamma$-ray \cite{J13,J14,J11} or a proton/nucleus \cite{J15,J16,J17,J18}) creates secondary photons and electrons which form the intergalactic electromagnetic cascade. If the extragalactic magnetic field (EGMF) is weak enough, then cascade electrons are not strongly deflected or delayed by it, and thus can contribute to the observable spectrum of a point-like source.

\begin{figure}
\centering
\includegraphics[width=8.0cm]{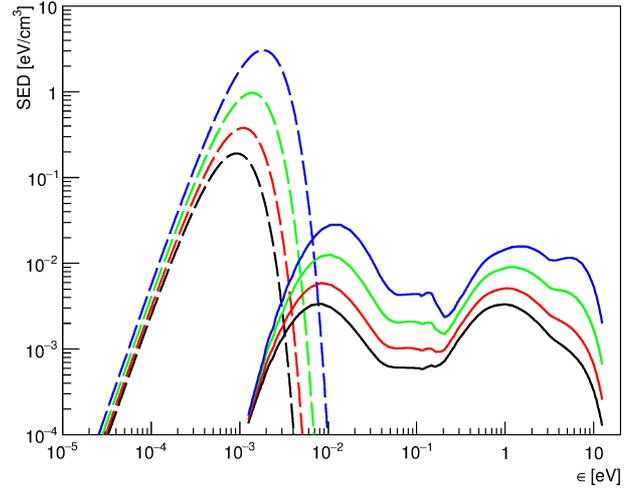}
\caption{SED of low-energy photon field of the Universe: the EBL (solid curves) and the CMB (dashed curves). Black curves: $z$= 0, red --- $z$= 0.186, green --- $z$= 0.5, blue --- $z$= 1.0.}
\label{fig-3}       
\end{figure}

\begin{figure}
\centering
\includegraphics[width=8.0cm]{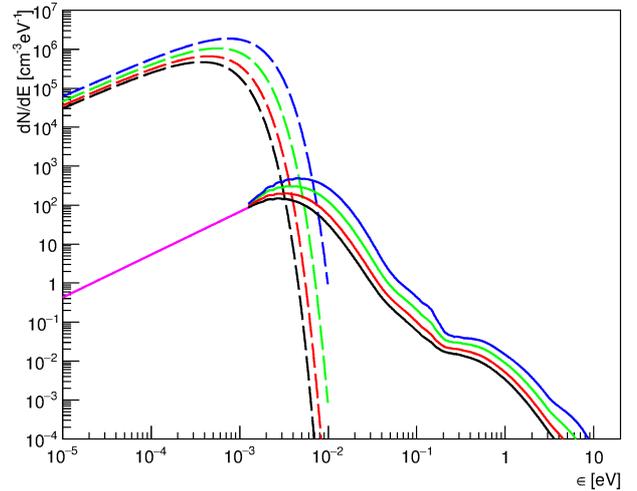}
\caption{Spectral density of the EBL and CMB for the same values of $z$ as in fig.~\ref{fig-3}. Magenta line shows an additional low-energy component that may be added to the model density of the EBL.}
\label{fig-4}       
\end{figure}

Electromagnetic cascade develops on a composite photon target consisting of a dense, but rather low-energy cosmic microwave background (CMB) and comparatively dilute extragalactic background light (EBL) which consists of IR, visible and UV photons. SEDs of these photon fields are shown in fig.~\ref{fig-3} for various redshift values assuming the EBL model of \cite{J19}. Starlight is the primary source of visible and UV EBL photons; the radiation of heated dust mainly forms the IR region of the EBL spectrum. For the case of moderate redshift $z<$1, primary $\gamma$-rays with energy $E_{0}<$50 $TeV$ produce pairs mainly on the EBL photons as the threshold energy of target photon for the case of the $\gamma\gamma\rightarrow e^{+}e^{-}$ process is about $\epsilon_{Thr}>$0.25 $ eV$/($E_{0}/$ $1 [TeV]$). On the other hand, most of the inverse Compton (IC) interaction acts occur on the CMB photons due to their high concentration and the absence of energy threshold for this process. Indeed, this issue is clear from fig.~\ref{fig-4} where the spectral density of the CMB and EBL vs. energy is shown. 

\begin{figure}
\centering
\includegraphics[width=8.0cm]{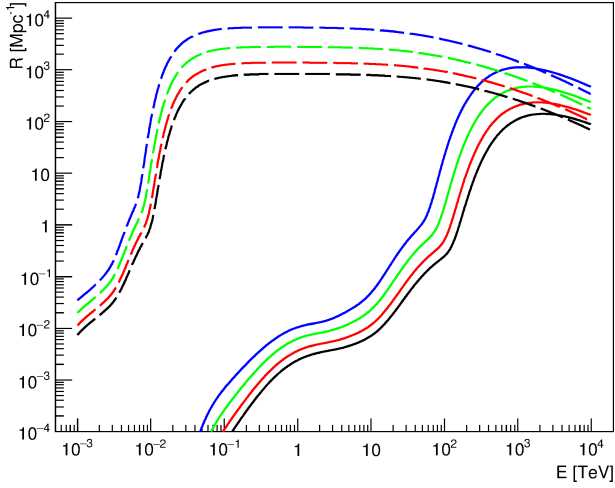}
\caption{Interaction rates for $\gamma$-ray and electron for the same values of $z$ as in fig.~\ref{fig-3}.}
\label{fig-5}       
\end{figure}

\begin{figure}
\centering
\includegraphics[width=8.0cm]{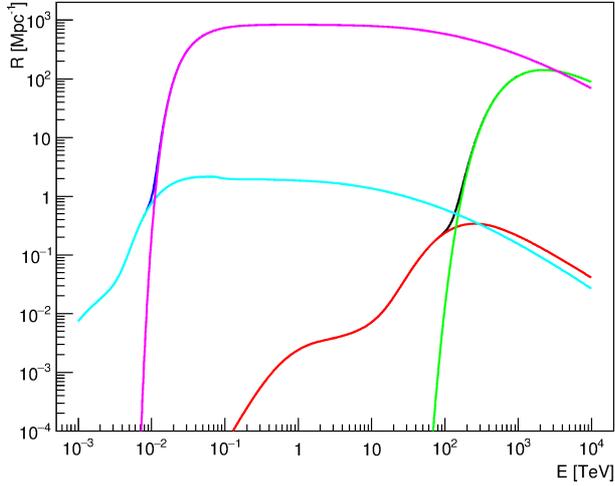}
\caption{Interaction rate for $\gamma$-ray and electron for $z$=0 decomposed on different components. Black line --- total $R_{\gamma}$, red --- $R_{\gamma}(EBL)$, green --- $R_{\gamma}(CMB)$, blue --- total $R_{e}$, cyan --- $R_{e}(EBL)$, magenta --- $R_{e}(CMB)$. Parameter $E_{thr}$ \cite{J20} was set to 3 $MeV$.}
\label{fig-6}       
\end{figure}

\begin{figure}
\centering
\includegraphics[width=8.0cm]{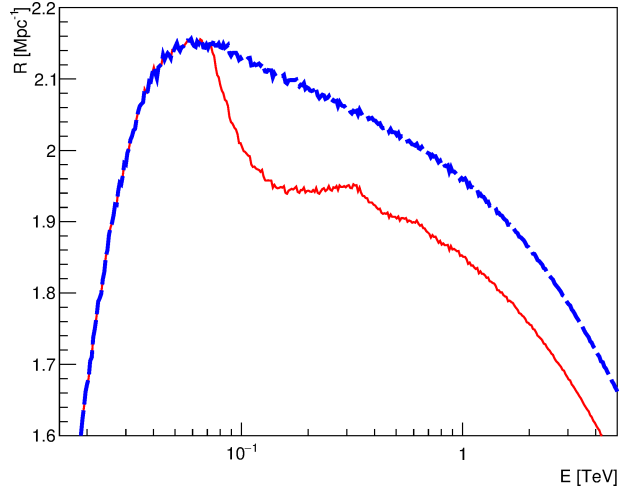}
\caption{Interaction rate $R_{e}(EBL)$ for $z$=0 (fragment) with (dashed thick blue curve) and without (solid red curve) additional component.}
\label{fig-7}       
\end{figure}

\begin{figure}
\centering
\includegraphics[width=8.0cm]{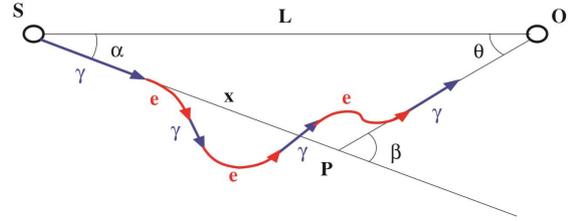}
\caption{A geometrical sketch of intergalactic electromagnetic cascade development (a figure from \cite{J23}). S denotes source, O - observer, $L$ is the distance from the source to the observer; $\beta$ is the total deflection angle of the last electron in the cascade chain, $\alpha$ --- the emission angle, $\theta$ --- the observation angle.}
\label{fig-8}       
\end{figure}

Following \cite{J20}, we calculated interaction rates $R(E,z)$ for $\gamma$-ray ($R_{\gamma}$) and $e$ ($R_{e}$) on the CMB and EBL for the energy range 1 $GeV$ -- 10 $PeV$= $10^{16}$ $eV$ (fig.~\ref{fig-5}). For $E>$ 100 $GeV$ $R_{e}(E/(1+z),z)/R_{e}(E,0)= (1+z)^{3}$ with good precision, reflecting the CMB evolution with $z$; this relation also approximately holds for the case of $\gamma$-ray and $E>$140 $TeV$; for $\gamma$-ray and $E<$100 $TeV$ \\ $R_{\gamma}(E/(1+z),z)/R_{\gamma}(E,0) \approx (1+z)^{p_{EBL}}$ with $p_{EBL}\approx$1.2--2.0 (depending on energy) due to the continuing stellar activity at $z<$1.

Fig.~\ref{fig-6} shows total interaction rates at $z$=0 for both $\gamma$-ray and $e$, as well as the partial contributions of the EBL and CMB to $R_{\gamma}$ and $R_{e}$. A small suppression in $R_{e}$ on the EBL at $E>$ 70 $GeV$ is induced by an abrupt cutoff of the model EBL spectrum at $\epsilon<10^{-3}$ $eV$. We show a comparison between $R_{e}(EBL)$ calculated with additional component denoted as magenta line in fig.~\ref{fig-4} and without such a component (see fig.~\ref{fig-7}). It is evident that without the artificial cutoff of the EBL spectral density at low energies the suppression in $R_{e}(EBL)$ is absent. While this spectral feature is clearly irrelevant for the final result on the observable spectrum of cascade $\gamma$-rays (its contribution to the total rate is $\sim$0.1 \%), it may appear of use while testing precision numerical Monte Carlo (MC) codes aimed at intergalactic cascade simulations.

The deflection of electrons in EGMF may impress certain spectral and timing signatures on the observable spectrum and time distribution of $\gamma$-rays (e.g. \cite{J21,J22}). Using an approximate geometrical scheme (from \cite{J23}, see fig.~\ref{fig-8}), \cite{J20} derived the following expression for the time delay of photons caused by the deflection of electrons in the EGMF ($c$ is the speed of light):
\begin{equation}
\Delta t \simeq (x(1+\sin\alpha/\sin\theta)-L)/c.
\end{equation}

Applying the small angle approximation ($\theta \simeq \sin\theta;  \cos\theta \simeq 1-\theta^{2}/2$) and the basic formula of trigonometry ($\sin\alpha = \sin(\beta-\theta)=\sin\beta\cos\theta - \cos\beta\sin\theta$) one can obtain that:
\begin{equation}
\Delta t \simeq \frac{x}{2c}\left(1-\frac{x}{L}\right)\sin^{2}\beta.
\end{equation}

Finally, we note that in the VHE energy range the additional broadening of cascade angular distribution in pair production and IC interaction acts and associated time delay are usually negligible (e.g. \cite{J24}).

\begin{figure}[!t]
\centering
\includegraphics[width=8.0cm]{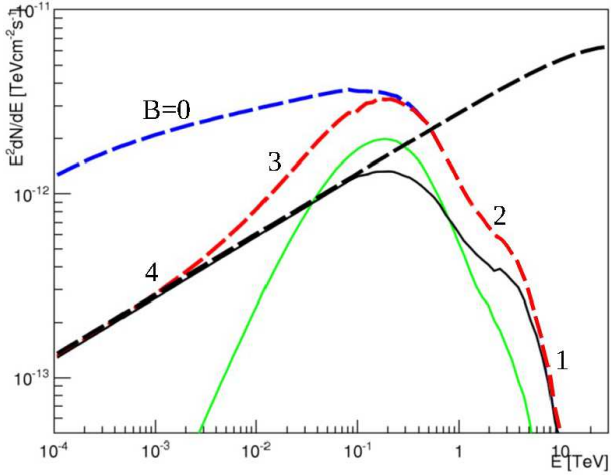}
\caption{Signatures of the electromagnetic cascade model ($z=0.186$) (a scheme from \cite{J25}). Green solid line denotes the cascade component, black solid line - primary absorbed component, black dashed line - primary intrinsic component, blue dashed line - observed spectrum with $B=0$, red dashed line - observed spectrum for the case of non-zero EGMF.}
\label{fig-9}       
\end{figure}

\section{Electromagnetic cascade model}
\label{sec-3}

Using the publicly available code ELMAG 2.02 \cite{J20}, we calculated a typical observable spectrum of a blazar with a hard intrinsic spectrum in the $TeV$ energy region (see fig.~\ref{fig-9}). In this picture, several spectral signatures \cite{J25} are clearly seen, namely: 1) a high-energy cutoff, 2) an ankle, which is due to the intersection of the primary component of $\gamma$-rays absorbed on the EBL and the cascade component, 3) a possible low-energy cutoff due to delay of cascade electrons (``magnetic cutoff''), 4) a ``second ankle'' at comparatively low energy where the primary component again starts to dominate over the cascade one. The first signature, the high-energy cutoff, is very similar to the case of the absorption-only model and thus is not specific for the electromagnetic cascade model. Provided that the cascade component is not entirely suppressed, the high-energy ankle is present in the spectrum irrespectively to the EGMF strength. On the other hand, the magnetic cutoff and the second ankle are connected with non-zero EGMF strength.

\begin{figure}[!t]
\centering
\includegraphics[width=8.0cm]{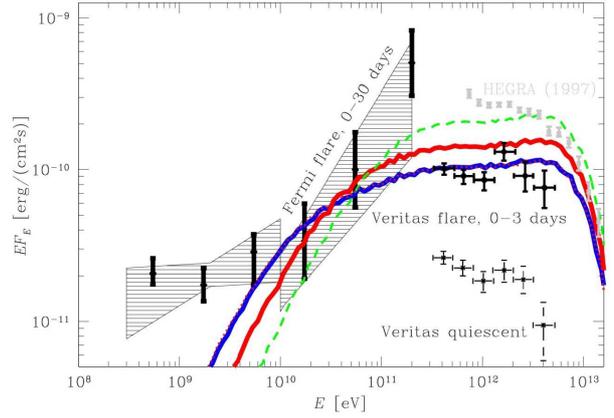}
\caption{Fermi LAT and VERITAS spectrum of Mrk 501 (2009 flare, a figure from \cite{J26}). Dots with errors denote the VERITAS measurements \cite{J27} of the spectrum during the first 3 days of the flare (solid lines) and in the quiescent state (dashed lines). Model curves show cascade spectra for $B$= $10^{-17}$ $G$ (green dashed), $3\cdot10^{-17}$ $G$ (red), $B = 10^{-16}$ $G$ (blue). The EGMF correlation length is 1 $Mpc$.}
\label{fig-10}       
\end{figure}

There are many indications that the cascade component indeed contributes to the observed spectra of some blazars at energies $E<$300 $GeV$. For instance, \cite{J26} reported the observation of a hard spectrum of Mkn 501 in the energy range 20-200 $GeV$. Fig.~\ref{fig-10} is a graph from this paper that shows the observed spectrum displaying two signatures of the electromagnetic cascade model (except the high-energy cutoff): a prominent magnetic cutoff and a possible ``second ankle''.

Other indication is the result of \cite{J28}, which hints at the larger number of blazars with comparatively hard spectra in the energy range 10-300 $GeV$ towards the directions to the Large Scale Structure (LSS) voids. These papers are discussed more thoroughly in \cite{J11}. In \cite{J29} an evidence for the existence of halos around various blazars was reported. The authors of this paper analyzed the spectra of 50 blazars. They divided them into two groups: BL Lac objects (24 sources) and Flat Spectrum Radio Quasars (26 sources). The first group had comparatively hard observed spectra and redshifts $z<0.5$. The second group sources could have any redshift and rather soft spectra. The analysis of these spectra showed that some blazars with relatively hard spectra could not be considered point sources. This result was interpreted in \cite{J29} as  evidence for the deflection of cascade electrons and positrons in the EGMF with the strength in the range of $10^{-17}-10^{-15}$ $G$.

However, all these indications can be explained by other phenomena that do not make use of intergalactic cascade development process. For instance, the authors of \cite{J26} proposed some alternative explanations to their findings in the paper itself. The significance of the effect observed in \cite{J28} is below 3 $\sigma$. As for \cite{J29}, the result could in principle be explained by the scattering of electrons and positrons accelerated in the source by a stronger local magnetic field ($B\sim10^{-12} - 10^{-7}$ $G$), in which case quasi-isotropic halos may form around the source \cite{J30}.

\begin{figure}
\centering
\includegraphics[width=8.0cm]{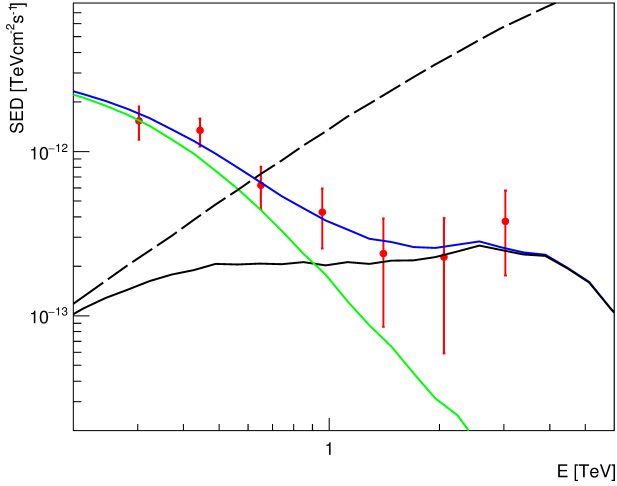}
\caption{Electromagnetic cascade model fit for 1ES 0347-121 for the case of $B$= 0 (a figure from \cite{J11}).}
\label{fig-11}      
\end{figure}

\begin{figure}
\centering
\includegraphics[width=8.0cm]{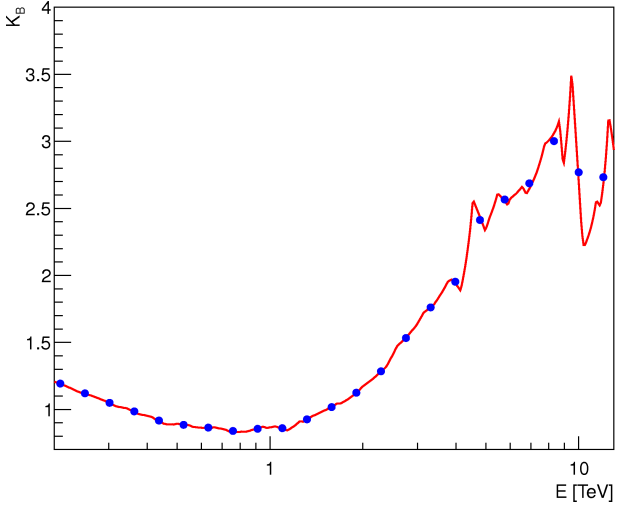}
\caption{Flux boost factor vs. energy for 1ES 0347-121 and the model with voidiness =1 (a figure from \cite{J11}).}
\label{fig-12}       
\end{figure}

\begin{figure}
\centering
\includegraphics[width=8.0cm]{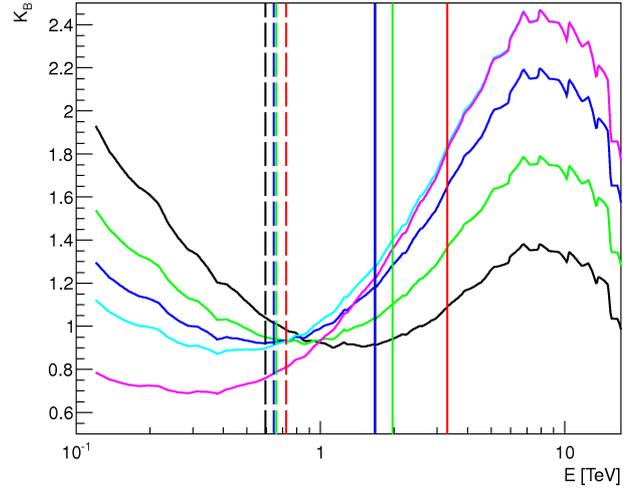}
\caption{Flux boost factor vs. energy for 1ES0229+200 and models with different voidiness $K_{V}$ (a figure from \cite{J11}): black curve --- $K_{V}$= 1, green curve --- 0.6, blue --- 0.4, cyan --- 0.3, magenta --- 0.2.}
\label{fig-13}       
\end{figure}

Let us consider the explanation of the anomaly \cite{J2,J3} in the framework of the electromagnetic cascade model (more details are available in \cite{J11}). Fig.~\ref{fig-11} presents a fit to the observed SED of blazar 1ES 0347-121 ($z$=0.188) assuming this model. Fig.~\ref{fig-12} shows the so-called ``flux boost factor'' --- the ratio of spectra for the case of the electromagnetic cascade model and the absorption-only model. At $E$=2-10 $TeV$ the value of $K_{B}$ clearly exceeds unity, therefore the electromagnetic cascade model allows to explain the observed excess in intensity in this energy region, with respect to the absorption-only model. Finally, fig.~\ref{fig-13} presents several curves of the boost factor vs. energy labeled by different values of the fraction of space filled with voids --- the so-called ``voidiness'' parameter \cite{J28}. It is interesting that at comparatively low values of voidiness (0.2-0.3) the values of $K_{B}$ are much larger at high energies than for the case of voidiness =1.

\section{Hadronic cascade model}
\label{sec-4}
\begin{figure}
\centering
\includegraphics[width=8.0cm]{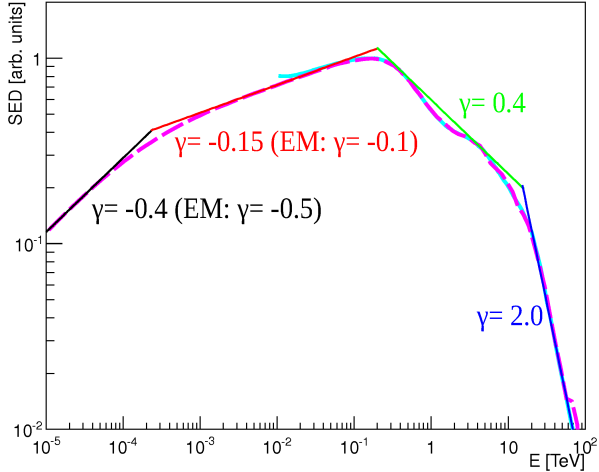}
\caption{Spectral signatures of the hadronic cascade model (primary proton energy $E_{p0}= 3\cdot10^{19} eV= 30  EeV$, $B=0$) (a figure from \cite{J11}). Straight solid lines denote different power indices that are present in the overall spectrum.}
\label{fig-14}       
\end{figure}

In hadronic cascade models, the primary particle is either a proton or a nucleus. They produce secondary particles on CMB and EBL via photopion processes and pair production; these secondary particles then initiate cascades the same way as they do in the electromagnetic cascade model. The graph of a typical observable SED of the hadronic cascade model for the case of zero EGMF and primary proton is shown in fig.~\ref{fig-14}. We approximated the spectrum with a broken poly-gonato power-law and compared the resulting power indices with the case of a purely electromagnetic cascade taken from \cite{J31}. Calculations of \cite{J31} apply to the case of the ``universal regime'', which usually sets for the case of sufficiently high primary energy ($E_{0}>$ 100 $TeV$) and sufficiently high redshift of the source.

The low-energy ($E<$ 200 $GeV$) indices for the hadronic model are nearly the same as for the electromagnetic cascade model (the latter are shown in parentheses, the former --- outside). However, at the higher energy the cutoff is not as marked for the case of the hadronic model. This fact makes it possible to effectively discriminate between EM and hadronic models and to identify the primary particle for the case of considered models. Cascade photons from nearby interactions of primary particles may form a significant excess over intensity expected for the case of the absorption-only model; therefore, the hadronic cascade model is, in principle, able to explain the VHE anomaly in blazar spectra \cite{J16,J17}.

In \cite{J11} we performed a detailed calculation of a large grid of hadronic cascade models with different parameters for the case of blazar 1ES 0229+200, assuming the emission model of \cite{J32}. It appeared that this model is plausible only for the case of comparatively low value of magnetic field circumventing the central emitting object, $B<$100 $nG$. Otherwise, the scattering of protons in this magnetic field significantly broadens the jet angular profile, thus weakening the total observable intensity in the VHE energy range much below the observed values.

\section{Conclusions}
\label{sec-5}

The first statistically significant (>3 $\sigma$) deviation from the absorption-only model was the VHE anomaly, that was initially interpreted as an indication for the $\gamma\rightarrow ALP$ oscillations. However, this interpretation is not supported by recent research; in particular, another important signature of $\gamma-ALP$ mixing --- the spectral irregularity at low energies --- was not found. Therefore, we considered two alternative models that could explain such an effect, namely, the electromagnetic and hadronic cascade models.

We briefly reviewed the phenomenology of electromagnetic cascade development on the CMB/EBL and discussed the main spectral signatures of a cascade component in the observable spectrum. Several works support the presence of the cascade component in the spectra of some blazars. As we have shown, a characteristic ``ankle'' signature at the intersection of the primary and secondary components may qualitatively explain the VHE anomaly. Finally, we briefly discussed the hadronic cascade model that  may also explain the anomaly, but demands a rather low strength of magnetic field circumventing the source.

To conclude, we have shown that several known deviations from the absorption-only model are naturally explained in the framework of the electromagnetic cascade model.

\section*{Acknowledgements}
The work was supported by the Russian Foundation for Basic Research (grant 16-32-00823). Figures 3-7 were produced during the visit of T.D. to the Tokyo University, Institute for Cosmic Ray Research (ICRR). For this part of work, T.D. acknowledges the support of the Students and Researchers Exchange Program in Sciences (STEPS) and the hospitality of the Tokyo University ICRR.

\end{document}